\documentclass[]{emulateapj}

\shorttitle{Molecular hydrogen deficiency in HI poor galaxies}
\shortauthors{Fumagalli et al.}

\begin{document}

\title{Molecular hydrogen deficiency in HI--poor galaxies and its implications for star formation}

\author{Michele Fumagalli \altaffilmark{1}, Mark R. Krumholz \altaffilmark{1}, J. Xavier Prochaska \altaffilmark{1}\altaffilmark{,2}, Giuseppe Gavazzi \altaffilmark{3} and Alessandro Boselli\altaffilmark{4}}
\email{mfumagalli@ucolick.org}

\altaffiltext{1}{Department of Astronomy and Astrophysics, University of California,  1156 High Street, Santa Cruz, CA 95064.}
\altaffiltext{2}{UCO/Lick Observatory; University of California, 1156 High Street, Santa Cruz,
CA 95064.}
\altaffiltext{3}{Departimento di Fisica G. Occhialini, Universita' di Milano--Bicocca, Piazza della scienza 3, Milano, Italy.}
\altaffiltext{4}{Laboratoire d'Astrophysique de Marseille, UMR 6110 CNRS, 38 rue F. Joliot-Curie, F-13388, Marseille, France.}

\begin{abstract}

We use a sample of 47 homogeneous and high sensitivity CO images taken from the Nobeyama and BIMA surveys to demonstrate that, contrary to common belief, a significant number ($\sim 40\%$) of HI--deficient nearby spiral galaxies are also depleted in molecular hydrogen. 
While HI--deficiency by itself is not a sufficient condition for molecular gas depletion,
we find that H$_2$ reduction is associated with the removal of HI inside the galaxy optical disk.
Those HI--deficient galaxies with normal H$_2$ content have lost HI mainly from outside their 
optical disks, where the H$_2$ content is low in all galaxies. This finding is consistent with theoretical models in which the molecular fraction in a galaxy is determined primarily by its gas column density. Our result is supported by indirect evidence that molecular deficient galaxies form stars at a lower rate or have dimmer far infrared fluxes than gas rich galaxies, as expected if the star formation rate is determined by the molecular hydrogen content. Our result is consistent with a scenario in which, when the atomic gas column density is lowered inside the optical disk below the critical value required to form molecular hydrogen and stars, spirals become quiescent and passive evolving systems. We speculate that this process would act on the time--scale set by the gas depletion rate and might be a first step for the transition between the blue and red sequence observed in the color--magnitude diagram. 
\end{abstract}


\section{Introduction}\label{intro}

\begin{deluxetable*}{l r r r l c c r r c l}
%\rotate
\tablewidth{0pt}
\tablecaption{Properties of our galaxy sample.}
\tablehead{
\colhead{Name}&\colhead{R.A.}&\colhead{Dec.}&\colhead{Dist.}&\colhead{Type}&\colhead{$L_{\rm H}$}&\colhead{R$_{25}$}&\colhead{$def_{\rm H_2}$}&\colhead{$def_{\rm HI}$}&\colhead{$H\alpha+[NII]$}&\colhead{CO Ref.}\tablenotemark{a}\\
\colhead{}&\colhead{(deg.)}&\colhead{(deg.)}&\colhead{(Mpc)}&\colhead{}&\colhead{(L$_\odot$)}&\colhead{('')}&\colhead{}&\colhead{}&\colhead{(E.W. \AA)}&\colhead{}} 
\startdata
IC 342   &  56.70   &  68.10  &  3.9 & Sc  & 10.5 & 598& -0.24 & -0.78&  56 &N \\
Maffei 2 &  40.48   &  59.60  &  3.4 & Sbc & 10.1 & 174& -0.10 & -0.17&   - &N \\
NGC 0253 &  11.89   & -25.29  &  3.0 & Sc  & 10.6 & 789&  0.10 &  0.65&  16 &N \\
NGC 0628 &  24.17   &  15.78  &  9.7 & Sc  & 10.5 & 293&  0.19 & -0.01&  35 &B$^*$ \\
NGC 1068 &  40.67   &  -0.01  & 14.4 & Sb  & 11.1 & 184& -0.11 &  0.47&  50 &N \\
NGC 2903 &  143.04  &  21.50  &  6.3 & Sb  & 10.4 & 352&  0.02 &  0.33&  18 &N$^*$ \\
NGC 3184 &  154.57  &  41.42  &  8.7 & Sc  & 10.2 & 222&  0.11 &  0.32&   - &N$^*$ \\
NGC 3351 &  160.99  &  11.70  &  8.1 & Sb  & 10.4 & 217&  0.34 &  0.60&  16 &N$^*$ \\
NGC 3504 &  165.80  &  27.97  & 26.5 & Sab & 10.7 &  73&  0.10 &  0.37&  35 &N \\
NGC 3521 &  166.45  &  -0.03  &  7.2 & Sb  & 10.6 & 249&  0.19 &  0.06&  20 &N$^*$ \\
NGC 3627 &  170.06  &  12.99  &  6.6 & Sb  & 10.5 & 306& -0.11 &  0.81&  19 &N$^*$ \\
NGC 3631 &  170.26  &  53.17  & 21.6 & Sc  & 10.7 & 111&  0.01 & -0.03&  34 &N \\
NGC 3938 &  178.20  &  44.12  & 17.0 & Sc  & 10.5 & 106& -0.06 & -0.24&  26 &B \\
NGC 4051 &  180.79  &  44.53  & 17.0 & Sb  & 10.6 & 146& -0.03 &  0.26&  32 &N \\
NGC 4102 &  181.60  &  52.71  & 17.0 & Sb  & 10.5 &  88& -0.02 &  0.47&  26 &N \\
NGC 4192 &  183.45  &  14.90  & 17.0 & Sb  & 11.0 & 293&  0.59 &  0.36&   9 &N \\
NGC 4212 &  183.91  &  13.90  & 16.8 & Sc  & 10.3 & 108&  0.17 &  0.57&  20 &N \\
NGC 4254 &  184.71  &  14.42  & 17.0 & Sc  & 10.9 & 184& -0.20 &  0.18&  32 &N \\
NGC 4258 &  184.74  &  47.30  &  6.8 & Sb  & 10.7 & 545&  0.44 &  0.25&  15 &B \\
NGC 4303 &  185.48  &	4.47  & 17.0 & Sc  & 11.0 & 197& -0.08 &  0.13&  36 &N \\
NGC 4321 &  185.73  &  15.82  & 17.0 & Sc  & 11.1 & 273&  0.05 &  0.53&  18 &N$^*$ \\
NGC 4402 &  186.53  &  13.11  & 17.0 & Sc  & 10.4 & 118&  0.02 &  0.73&  12 &N \\
NGC 4414 &  186.61  &  31.22  & 17.0 & Sc  & 10.9 & 108& -0.09 & -0.15&  23 &N \\
NGC 4419 &  186.74  &  15.05  & 17.0 & Sa  & 10.6 & 105&  0.40 &  1.04&   7 &N$^*$ \\
NGC 4501 &  188.00  &  14.42  & 17.0 & Sbc & 11.2 & 216&  0.14 &  0.51&   6 &N \\
NGC 4535 &  188.58  &	8.20  & 17.0 & Sc  & 10.8 & 249&  0.02 &  0.35&  17 &N \\
NGC 4536 &  188.61  &	2.19  & 17.0 & Sc  & 10.7 & 216&  0.18 &  0.29&  20 &N$^*$ \\
NGC 4548 &  188.86  &  14.50  & 17.0 & Sb  & 10.9 & 180&  0.84 &  0.70&  17 &N \\
NGC 4569 &  189.21  &  13.16  & 17.0 & Sab & 11.0 & 321&  0.23 &  1.13&   2 &N$^*$ \\
NGC 4579 &  189.43  &  11.82  & 17.0 & Sab & 11.1 & 188&  0.62 &  0.71&   4 &N \\
NGC 4654 &  190.99  &  13.13  & 17.0 & Sc  & 10.7 & 149& -0.02 &  0.17&  31 &N$^*$ \\
NGC 4689 &  191.94  &  13.76  & 17.0 & Sc  & 10.5 & 175&  0.14 &  1.06&  14 &N$^*$ \\
NGC 4736 &  192.72  &  41.12  &  4.3 & Sab & 10.4 & 232&  0.78 &  0.68&  10 &N$^*$ \\
NGC 4826 &  194.18  &  21.68  &  4.1 & Sab & 10.3 & 314&  0.70 &  0.86&   9 &B$^*$ \\
NGC 5005 &  197.73  &  37.06  & 21.3 & Sb  & 11.3 & 143&  0.28 &  0.47&   7 &B \\
NGC 5033 &  198.36  &  36.59  & 18.7 & Sc  & 11.0 & 293& -0.16 &  0.19&  20 &B \\
NGC 5055 &  198.96  &  42.03  &  7.2 & Sbc & 10.7 & 352&  0.04 &  0.12&  20 &N$^*$ \\
NGC 5194 &  202.47  &  47.19  &  7.7 & Sbc & 10.8 & 232& -0.36 &  0.12&  28 &N$^*$ \\
NGC 5236 &  204.25  & -29.86  &  4.7 & Sc  & 10.7 & 464&  0.23 & -0.49&  33 &N$^*$ \\
NGC 5247 &  204.51  & -17.88  & 22.2 & Sb  & 10.9 & 161&  0.06 &  0.05&   - &N \\
NGC 5248 &  204.38  &	8.88  & 22.7 & Sb  & 11.1 & 122& -0.10 & -0.25&  26 &N \\
NGC 5457 &  210.80  &  54.35  &  5.4 & Sc  & 10.5 & 719&  0.06 &  0.28&  31 &N$^*$ \\
NGC 6217 &  248.16  &  78.20  & 23.9 & Sbc & 10.5 &  67&  0.13 & -0.45&  34 &N \\
NGC 6946 &  308.72  &  60.15  &  5.5 & Sc  & 10.4 & 344& -0.28 & -0.45&  34 &N$^*$\\
NGC 6951 &  309.31  &  66.10  & 20.3 & Sb  & 11.1 &  97&  0.24 &  0.04&  16 &N\\
NGC 7331 &  339.27  &  34.42  & 14.3 & Sbc & 11.1 & 273&  0.14 & -0.04&   - &B$^*$ \\
UGC 2855 &   57.09  &  70.13  & 20.3 & Sc  & 10.9 & 132& -0.34 & -0.13&  34 &N \\
\enddata
\tablenotetext{a}{N: Nobeyama survey \citep{kun07}; B: BIMA survey \citep{hal03}. Galaxies 
included in the high--resolution study are identified by an asterisk.}
\tablerefs{References are discussed throughout the text.}\label{tab1}
\end{deluxetable*}

A complete and coherent description of the detailed connection between a galaxy's gaseous phase and its star formation is still lacking. This poses a significant limitation for studying galaxy evolution, even in the local universe. From an observational point of view, two different approaches have been adopted to address this fundamental topic. The first takes advantage of high spatial resolution images to compare directly the molecular and atomic phases of the interstellar medium (ISM) with recent star formation activities. The results \citep {won02,ken07,fum08,big08} suggest that stars form preferentially in molecular regions, as traced by diffuse CO emission. The second approach is more indirect,
but relies on estimators for the star formation rate and atomic gas that can be easily collected for large samples. These projects statistically study the effects of the atomic gas removal on star formation in cluster galaxies compared with field galaxies. 
Results \citep[][Gavazzi et. al, in prep.]{ken83,gav02,koo04,gal08} show a significant correlation between the atomic gas and the star formation: the latter is quenched in those galaxies which are HI deficient due to environmental effects. Although this result can be interpreted in terms of HI feeding star formation \citep{bos01}, it is possible that  molecular hydrogen (H$_2$) depletion also plays a critical role.

However, observations to date suggest that molecular hydrogen, which dominates the mass of the 
interstellar medium  in the center of spiral galaxies, is not strongly affected by environmental perturbations. Galaxies in rich clusters appear to have on average the same molecular content
as field galaxies \citep{sta86,ken86,ken89,cas91,bos97,bos02}. This result is consistent with a simple theoretical argument based on the assumption  that ram-pressure stripping is the dominant environmental process,  a fact that, although not yet proved, is likely to be the correct interpretation for HI observations in the Virgo and Coma clusters \citep[see a review by][]{bos06}.  According to this picture,  a galaxy which travels at a velocity $v_{\rm rel}$ relative to the dense intergalactic medium (IGM) or intracluster medium (ICM) characterised by a density $\rho_{\rm IGM}$  looses its ISM at the escape velocity $v_{\rm esc}$ and density $\rho_{\rm ISM}$ only if 
\begin{equation}
\rho_{\rm IGM} v_{\rm rel}^2 > \rho_{\rm ISM} v_{esc}^2\:.
\end{equation}    
This condition is not easily satisfied for gas in molecular hydrogen phase, which has densities up to 10$^5$ cm$^{-3}$, and is usually strongly bound to the galaxy due to its proximity to the galactic center.

This theoretical argument seems to agree with observations of the CO abundance in members of the Virgo and Coma clusters. Using a sample composed of 47 Virgo  galaxies, \citet{sta86} showed that the molecular fraction CO/HI increases towards the center of the cluster, identified by the position of M87. This is consistent with the removal of atomic hydrogen only from inner cluster galaxies, without any corresponding change in the molecular hydrogen content. With independent 
CO measurements, the same conclusion was reached by \citet{ken86,ken89},  who studied molecular hydrogen deficiency in datasets  of  23 and 41 Virgo galaxies. Their analysis shows that the molecular content has not responded to the HI removal on time scales of $10^9$ yr, i.e. it is unperturbed within the Virgo crossing time. Furthermore, \citet{cas91} pointed out that even in the Coma cluster, where the ram--pressure is stronger than in Virgo, there is no correlation between the H$_2$ mass and the atomic hydrogen content of individual spirals. Surprisingly, using the data collected by \citet{ken89} with updated distances, \citet{ren92} found from a comparison of normalized H$_2$ masses that some deficiency of molecular hydrogen holds among Virgo galaxies, as compared to the field. This result is in contrast with a series of papers by \citet{bos97,bos02}, who did not find any correlation between the HI and the H$_2$--deficiency on a large sample of $\sim 250$ Virgo and Coma galaxies.
Except some extreme cases of galaxies which have their molecular gas depleted and displaced \citep[e.g.][]{vol08}, there is no strong evidence that cluster member galaxies have statistically
less molecular gas than field galaxies. 

This seems in contradiction with a coherent picture of the star formation in spirals.
If stars really do form primarily in molecular gas, one might expect the quenching in the star formation activity to be connected with a molecular gas deficiency in HI poor galaxies. This hypothesis is partially supported by \citet{fum08}, who, using a small sample of 28 images from the Nobeyama CO survey \citep{kun07}, find some hints of a possible depletion in the molecular content (see their Fig. 4 and 6). The aim of this paper is  test for H$_2$--deficiency by using homogeneous and high sensitivity data from the Nobeyama \citep{kun07}, BIMA \citep{hal03}, VIVA (Chung at al., in prep.) and THINGS \citep{wal08} surveys.
At first, we compare the integrated molecular and atomic gas mass; then, we study the distribution of the gas as a function of the galaxy radius and finally we compare both the atomic and hydrogen gas content with indicators for the star formation. 
While previous studies selected galaxies mainly from the Virgo and Coma clusters, our sample includes also objects from nearby groups. In these diverse environments it is less obvious that ram--pressure is the most efficient process in removing the atomic gas. Therefore, in our analysis we will focus only on the effects
produced by the HI removal, without addressing to the details of its origin that is likely connected 
with a generic environmental perturbations or galaxy feedback.

The paper is organized as follows: the sample and the data reduction are discussed in Sec. \ref{datared}, the analysis is given in Sec. \ref{analysis} and the discussion in Sec. \ref{discussion}. In particular, we focus on why previous surveys did not detect the H$_2$--deficiency (\S \ref{subdisc1}), we explain the process responsible for the molecular gas depletion (\S \ref{subdisc2}) and we discuss the implications of H$_2$--deficiency on the star formation (\S \ref{subdisc3}). The conclusions are in Sec. \ref{conclusion}.

\section{Data Reduction}\label{datared}

\subsection{The sample}
Our sample is selected based on the criterion of high sensitivity CO images
and homogeneity to minimize the scatter in CO measurements.   
Because previous CO surveys show a relative flux dispersion which severely limits the H$_2$ deficiency analysis (see \S \ref{subdisc1}), we limit to the Nobeyama CO survey \citep{kun07} and the BIMA survey \citep{hal03} which are shown to agree to within 0.4 dex. 47 massive spiral galaxies in the H--band luminosity range $10^{10} \leq L_{\rm H}/L_\odot \leq 10^{11}$ meet our selection criterion, 40 from the Nobeyama and 7 from BIMA. Our sample includes only spiral galaxies since early--type (E/dE) galaxies are known to be extremely gas poor \citep{dis07}, while late--type galaxies (Sa--Sc, BCD or Irr/dIrr) are typically gas rich \citep{gav08}. 
Table \ref{tab1} summarizes some observational properties for our sample.
Since we are dealing mainly with the full Nobeyama CO atlas, which has been selected using a far infrared (FIR) emission criterion ($F_{100\mu m}>10$ Jy), our sample may be biased against CO poor galaxies. 

The Nobeyama maps are obtained with a 45m single--dish radio telescope that provides high sensitivity ($\sim$ 0.5 M$_\odot$ pc$^2$) flux measurements for the entire galaxy disk. The spatial resolution is about $15''$, which corresponds to an average physical scale of $\sim1$ kpc at the mean distance of our sample ($\sim15$ Mpc). In contrast, despite its higher spatial resolution ($6''-10''$, corresponding to 450--700 pc), the interferometric BIMA survey has systematically poorer sensitivity to extended flux because  interferometers cannot  detect flux from regions larger than the size--scale set by the minimum distance between the antennas. One can recover the total flux by using single dish observations that provide the zero spacing element. We have limited  our galaxy sample to the subset with single--dish data.  Even for these galaxies,  the  reconstruction algorithm is biased against large scales \citep[][app. A]{hal03}; however, a comparison
between Nobeyama and BIMA data indicates that the missed flux is negligible compared to the total flux, i.e.
the BIMA survey provides a reliable estimate of the integrated molecular hydrogen mass. When a galaxy is imaged in both the surveys, we use the Nobeyama data.

For our analysis we will also compare the spatial distributions of molecular and atomic hydrogen; therefore we collect spatially resolved HI images from the THINGS survey \citep{wal08} and the VIVA survey (Chung et al., in prep.). Since spatially resolved images are not available for all the galaxies in our sample, while studying  surface density profiles we restrict our analysis to a subsample of 20 galaxies (see Table \ref{tab1}).

\subsection{A multifrequency analysis}
The comparison between the molecular distribution, the atomic hydrogen content and the star formation activity
requires some preliminary transformations from observed to physical quantities. In this section we describe this issue.

\subsubsection{The molecular hydrogen}
The H$_2$ molecule has no permanent dipole so we must infer its abundance indirectly by assuming that the conditions of the ISM where the H$_2$ forms are similar to those in which other molecules are synthesized. A good tracer for H$_2$ is CO because the intensities of its rotational transitions are correlated to the 
molecular column density. This is based on a simple argument which relates the velocity inferred from the line profile with the virial velocity of molecular clouds \citep{dic86,sol87}.

The Nobeyama and BIMA surveys image the CO $J:1-0$ transition at 115 GHz, which is excited above 5.5 K in cool gas at moderate density. We  infer the molecular hydrogen column density by adopting a conversion factor $X$ (cm$^{-2}$ (K km/s)$^{-1}$), which is a function of temperature, density, UV radiation field, metallicity and even of the shape of the molecular clouds \citep[][app. A]{wal07,bos02,bos97}. 
Because it is impossible to model $X$ theoretically for actual galaxies, a constant conversion factor 
derived using observations in our Galaxy has often been assumed \citep[e.g.][]{sol87}. The Galactic factor likely gives underestimates for the molecular hydrogen content in low mass galaxies, because empirical calibrations show that $X$ decreases for increasing mass \citep[e.g.][]{bos02}. 
Since we expect the CO--to--H$_2$ conversion factor to depend primarily on the dust and metallicity that regulate the formation and photodissociation of H$_2$ and the CO heating, we choose an individual $X$ value for each galaxy following the empirical calibrations of \citet{bos02}.
This is
\begin{equation}\label{Xbos}
\log X =-0.38 \log L_{\rm H} +24.23\:,
\end{equation}
where $L_{\rm H}$ is the H--band luminosity in L$_\odot$. We retrieve $L_{\rm H}$ from GOLDMine \citep{gav03} and SIMBAD. 
This calibration accounts for the dependence of $X$ on the metal abundance because 
the H--band luminosity correlates with galaxy metallicity \citep[e.g.][]{bos02}.
The systematic uncertainty on the $X$ factor derived using eq. (\ref{Xbos}) ranges from 0.65 to 0.70 dex
for the luminosity interval $10^{10}-10^{11}$ L$_\odot$. Other unknown dependencies can increase this error. The final conversion for the molecular hydrogen surface density is:
\begin{equation}
\Sigma_{H_2}\:[{\rm M_\odot\:cm^{-2}}]=2 m_p  X F_{\rm co}\:,
\end{equation}	   
where $m_p$ is the proton mass in M$_\odot$ and $F_{\rm co}$ is the CO flux in K km/s. This conversion does not include the mass of helium, but its contribution may be included trivially by multiplying the derived surface density by 1.36. The integrated CO flux ($\bar F_{\rm co}$ in Jy km/s) is converted into a total H$_2$ mass ($M_{\rm H_2}$ in M$_\odot$) using \citep{hal03}:
\begin{equation}
M_{H_2}=3.92\times 10^{-17} X \bar F_{\rm co} D^2\:,
\end{equation}	   
where $D$ is the galaxy distance in Mpc from GOLDMine and \citet{tul94}.

\subsubsection{The atomic hydrogen}\label{HIdata}
We estimate the atomic hydrogen content via 21 cm emission line.
Single--dish HI surveys, including the ALFALFA \citep{gio05} or HIPASS \citep{mey04} surveys, do provide the HI integrated flux for all the galaxies in our sample. Collecting different HI data for a same galaxy, we notice a significant dispersion in the measured HI flux, even up to a factor of 2. Because we are interested in a relative comparison between galaxies rather than an accurate determination of the absolute HI mass, we retrieve all the HI fluxes from the HyperLeda database \citep{pat03}. Even if more recent measurements have  better sensitivity and provide a more accurate determination of the total HI abundance, this approach gives an homogeneous dataset \citep{pat03II}. As discussed in Appendix \ref{appHIfl}, a particular choice for the observed flux does not affect our analysis. 

The integrated HI flux ($\bar F_{\rm HI}$ in Jy km/s) is converted into a total mass ($M_{\rm HI}$ in M$_\odot$) using
\begin{equation}
M_{\rm HI}=2.36\times10^5 \bar F_{\rm HI} D^2\:,
\end{equation}
where $D$ is the galaxy distance in Mpc from GOLDMine and \citet{tul94}.
The HI surface density profile is computed from THINGS and VIVA images according to the conversion
\begin{equation}
\Sigma_{HI}\:[{\rm M_\odot\:cm^{-2}}]=m_p N_{\rm HI}\:,
\end{equation}	   
where $N_{\rm HI}$ is the column density in cm$^{-2}$ derived from the flux $F_{\rm HI}$ (Jy/beam km/s) using 
the following calibration \citep[e.g.][]{wal08}:
\begin{equation}
N_{\rm HI}[{\rm 10^{19} cm^{−2}}]=110.4\times 10^3  \frac{F_{\rm HI}}{B_{\rm min}\times B_{\rm max}}
\end{equation} 
with $B$ the beam FWHM in arcsec. As with H$_2$, these surface densities do not include the mass of helium.

\subsubsection{The star formation activity}
In the ISM of galaxies,  the $H\alpha$ emission line is caused by the recombination of atomic hydrogen which has been ionized by the UV radiation of O and B stars. Therefore $H\alpha$ is a tracer of recent, massive ($M>10$ M$_{\odot}$) star formation activity on time scales of $t\leq 4\times 10^6$ yr \citep{ken98}. For all but three objects, we retrieve from the literature \citep[GOLDmine;][]{ken83,jam04,ken08} the $H\alpha$ equivalent width (E.W.), a tracer of the specific star formation rate \citep{ken83II}. 

Since the narrow band filter used to image the $H\alpha$ emission line is large enough to collect the flux from the $[NII]$ line, the measured E.W. overestimates the $H\alpha$ emission. Unfortunately, the ratio $[NII]/H\alpha$ is not available for all of the galaxies in our sample and we can apply only the average correction $[NII]/H\alpha = 0.53$, as suggested by \citet{ken92}. However, this value underestimates the $[NII]$ emission from the bulges of galaxies that host active galactic nuclei (AGN), where the nuclear $[NII]$ can be even stronger than the $H\alpha$. In addition, even if the nuclear emission were corrected using a suitable factor, it would not be possible to disentangle between the $H\alpha$ flux due to star formation from that due to AGN ionization. Our sample, composed of massive spirals, is likely to be affected by this contamination \citep{dec07}. 

Another effect to consider is dust absorption. In principle one can estimate the extinction by measuring the $H\alpha/H\beta$ or the  $H\alpha/Br_\gamma$ fluxes (case B recombination). These values, however, 
are unknown for the majority of our galaxies. One could apply a standard dust extinction  \citep[e.g. $A=1.1$ mag,][]{ken83} but this would not change the relative distribution of the measurements. Instead,  we simply keep the measured $H\alpha+[NII]$ E.W. as an approximation for the specific star formation rate. The reader should keep in mind that  a large uncertainty is associated with these measurements, in addition to the error in the flux determination. 

\subsection{Global fluxes and surface brightness profiles}
We compute total CO fluxes from the Nobeyama images using the IRAF task {\tt qphot} by integrating the emission in an aperture which encloses the entire galaxy.
Fluxes from the BIMA survey are retrieved from \citet{hal03}, who list values computed within a square region.
We retrieve the HI flux from HyperLeda, as previously discussed. We compute the surface brightness profiles for the CO and HI using an IDL procedure that averages the flux into elliptical annuli characterised by three fixed parameters (the galaxy center, the eccentricity and the position angle); we adopt a step along the major axis of $\sim 5$ arcsec, roughly 1/3 of the spatial resolution of the CO images from the Nobeyama survey.
All the profiles are corrected for the projection effect using the galaxy inclination (HyperLeda) on the plane of the sky. For each galaxy we finally obtain the HI and H$_2$ surface density profiles 
$\Sigma_{HI}$ and $\Sigma_{H_2}$  in $\rm M_\odot\:pc^{-2}$.

\begin{figure*}
\plottwo{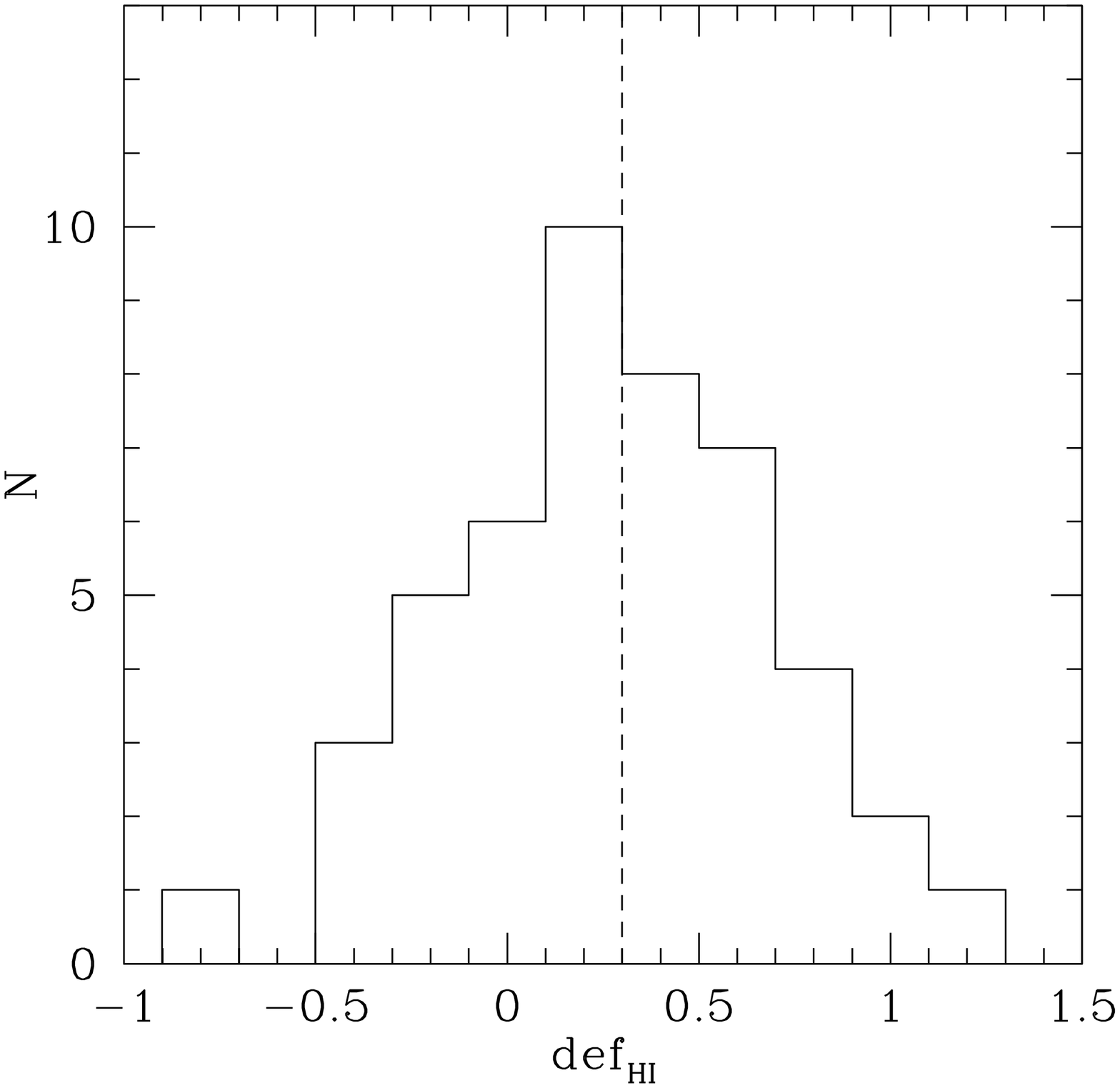}{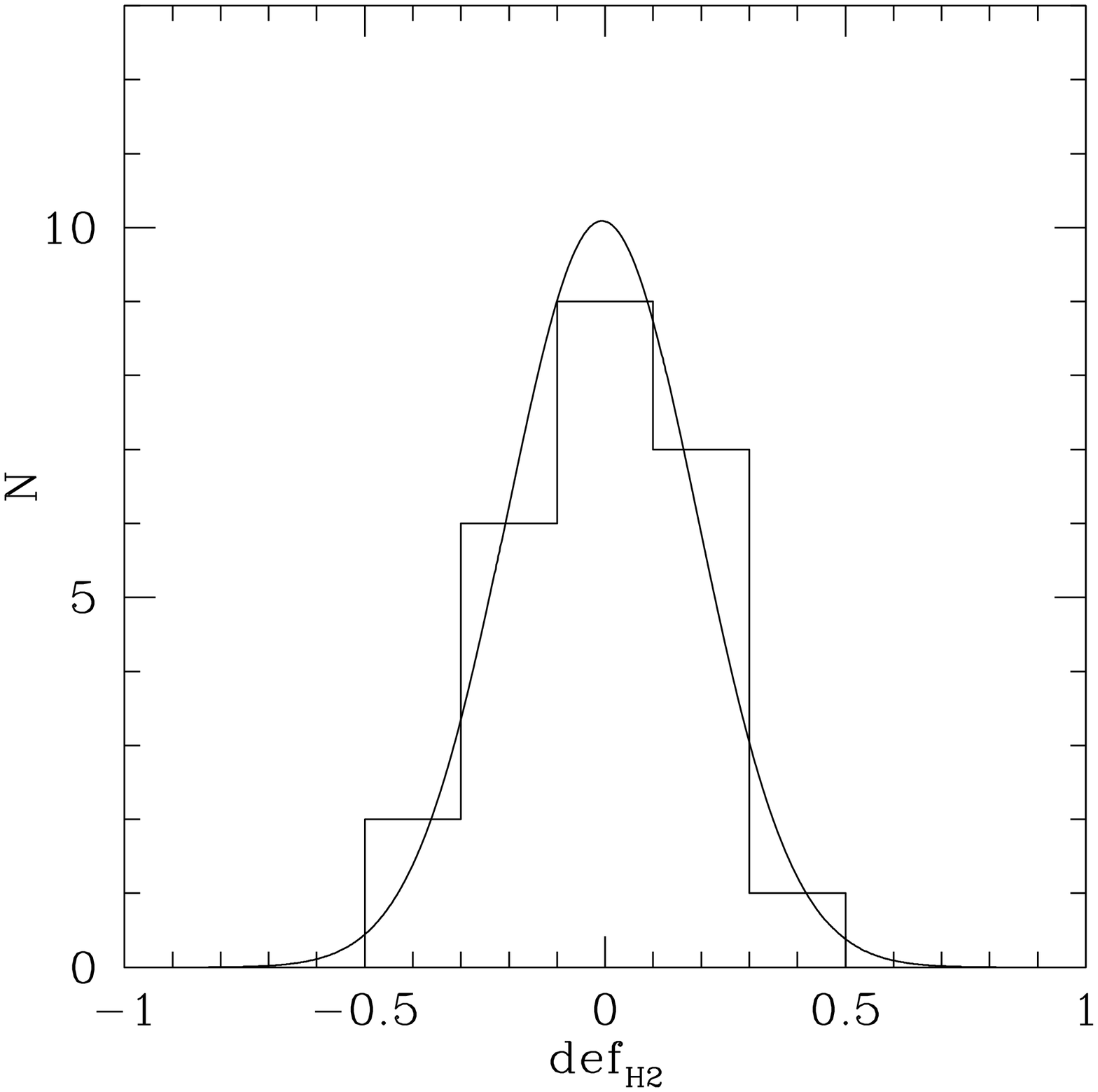}
\caption{a - Histogram of the HI--deficiency values in our sample, separated in isolated and perturbed galaxies at $def_{\rm HI}=0.3$. A positive value of $def_{\rm HI}$ indicates that a galaxy has an HI mass lower than expected on average according to its size and morphology. b - Histogram of the H$_2$--deficiency values for isolated galaxies only. The Gaussian fit shows that $\sigma=0.2$.}\label{histdef}
\end{figure*}

\section{Analysis}\label{analysis}

To test for molecular hydrogen depletion in our dataset, we compare the H$_2$ mass in a sample of isolated galaxies with that in perturbed galaxies.  Our isolation criterion is defined by the atomic hydrogen content. Because the edges of the HI disk are weakly bounded, any perturbation can easily remove atomic gas from a galaxy, lowering its HI mass. That the HI--deficiency should trace environmental effects is supported by \citet{gio85}, who show that HI--deficiency increases toward the center of a cluster. In these regions the IGM is dense, the cluster potential well is deep and the number of galaxies is high so that environmental processes are very effective. 
We expect HI deficiency to be a useful measure of disturbance even if environment is not the dominant means of gas removal. For example, if star formation feedback is the primary cause of gas loss, this too will preferentially remove the low--density, loosely bound HI.

Formally, the atomic gas depletion is quantified using $def_{\rm HI}$, the deficiency parameter, defined by
\citet{hay84} as the logarithmic difference between the expected HI mass in isolated galaxies and the observed value:
\begin{equation}
def_{\rm HI}=\log M_{\rm HI,exp} - \log M_{\rm HI,obs}
\end{equation}  
The reference mass  $M_{\rm HI,exp}$ is computed as a function of the optical radius at the 25th mag/arcsec$^2$ in B band (retrieved from NED and HyperLeda) using the coefficients recently updated by \citet{sol96},  which are weakly dependent on the morphological type.
Several sources of uncertainty affect the determination of $def_{\rm HI}$: the error on the observed HI mass, the statistical uncertainty on the calibrated coefficients and the determination of the optical diameter.
It is hard to quantify a value for the final uncertainty, but a comparison of the values derived for a single galaxy using different data available in the literature suggests a mean dispersion of $0.2-0.3$ dex on the HI--deficiency.

Figure  \ref{histdef}a shows the distribution of the HI--deficiencies for our sample. If we arbitrarily consider unperturbed those galaxies which have lost less than a factor of 2 in their HI mass ($def_{\rm HI}<0.3$), our subsample of isolated galaxies is composed of 25 objects with a mean deficiency  $def_{\rm HI}=-0.045\pm0.27$. This is consistent with zero. Moreover, we point out that the dispersion on the mean value is within the deficiency threshold 
of 0.3 dex. Only 4 galaxies lie at  $def_{\rm HI}<-0.3$, indicating that our reference sample is not dominated by galaxies with HI mass above the average. The remaining 22 galaxies  have $def_{\rm HI}\geq0.3$,  spanning  the entire range of moderate and high deficiencies ($0.3-1.2$ dex).

In a similar way,  we quantify a possible reduction in the molecular hydrogen content using the H$_2$--deficiency parameter $def_{\rm H_2}$ \citep{bos97,bos02}:
\begin{equation}
def_{\rm H_2}=\log M_{\rm H_2,exp} - \log M_{\rm H_2,obs}\:.
\end{equation}  
The value for the reference mass $M_{\rm H_2,exp}$ is computed from the observed correlation between the H$_2$ and the H--band luminosity \citep{bos97} for isolated galaxies ($def_{\rm HI}<0.3$). This definition is based on the natural scaling of the molecular gas abundance with the size of a galaxy, i.e. more massive spiral galaxies are more gas rich. Although our sample covers a small range in luminosity, owing to the good quality of our data (see section \ref{subdisc1}), we choose to compute a new reference mass with a least square fit: 
\begin{equation}\label{hlco}
\log M_{\rm H_2,exp}=0.50 \log L_{\rm H} + 3.98\:.
\end{equation}  
Since the dispersion (rms) in the calibration is $\sim  0.2$ (See also Fig. \ref{histdef}b), galaxies with $def_{\rm H_2}<0.2$ cannot be distinguished from normal galaxies. In the following analysis we therefore consider a galaxy H$_2$--deficient only if $def_{\rm H_2} \geq 0.2$. However, as we shall see below, $def_{\rm H_2}$ shows a continuous rather than a bimodal distribution. Thus, this criterion is arbitrary. It does have the virtue, though, of nicely dividing our sample into classes which have a comparable number of objects.
As with $def_{\rm HI}$, it is hard to quantify an uncertainty for $def_{\rm H_2}$. Deficiency is a relative quantity, so the level of uncertainty for  $def_{\rm H_2}$ is likely to be significantly lower than for the absolute H$_2$ mass. The error of 0.7 dex we quoted in equation  (\ref{Xbos}) is therefore an upper limit on the scatter in $def_{\rm H_2}$. A more reliable estimate of the error is the dispersion
of 0.2 dex we measure in  $def_{\rm H_2}$ for the control sample of isolated galaxies (Fig. \ref{histdef}b). Since random error in computing M$_{\rm H_2}$, for example galaxy--to--galaxy scatter in $X$, can only broaden the intrinsic dispersion of H$_2$ masses at fixed L$_H$, our measured scatter of 0.2 dex sets a more realistic upper limit on the size of random errors. The one caveat is that this value may neglect systematic uncertainties if unperturbed and perturbed galaxies have different values of luminosity or metallicity, for example, and thus systematically different values of $X$. However, we do not observe or expect any obvious trend of these quantities as a function of $def_{\rm HI}$ or $def_{\rm H_2}$.

\begin{figure}
\includegraphics[scale=.40]{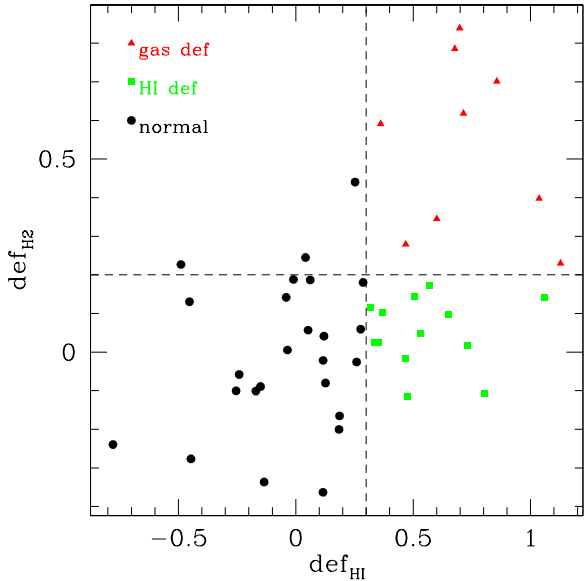}
\caption{Comparison between calculated $def_{\rm H_2}$ and $def_{\rm HI}$ values for our  sample. The vertical dashed line corresponds to $def_{\rm HI}=0.3$ dex, the value below which a galaxy is considered unperturbed; the horizontal line represents  $def_{H_2}=0.2$ dex. The absence of galaxies in the upper--left quadrant suggests that HI--deficiency is only a necessary condition for H$_2$--deficiency.}\label{codef}
\end{figure}

\begin{figure}
\includegraphics[scale=.40]{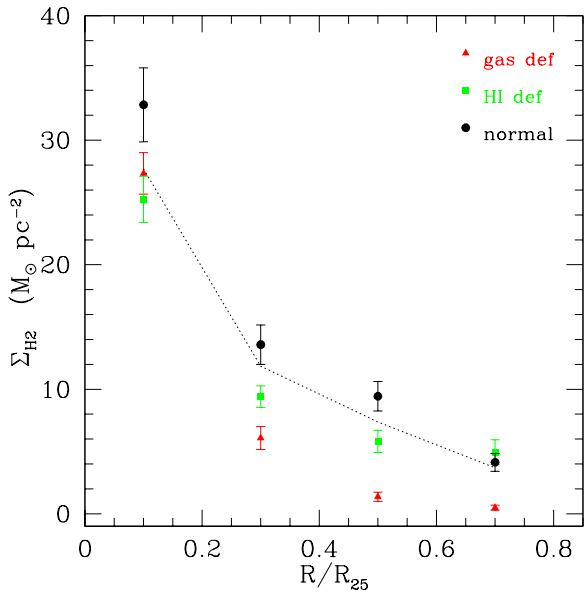}
\caption{Average H$_2$ profiles in bins of normalized radius. The dashed line
is computed by excluding HI--rich galaxies with $def_{\rm HI}<-0.3$. The offset among the profiles reflects the $def_{H_2}$ distribution in Fig. \ref{codef}.}\label{coprof}
\end{figure}

Figure \ref{codef} shows a comparison between  $def_{\rm H_2}$ and $def_{\rm HI}$ in our sample.
The vertical dashed line corresponds to $def_{\rm HI}=0.3$ dex, the value below which a galaxy is
considered unperturbed; the horizontal line represents  $def_{\rm H_2}=0.2$ dex.
Inspecting this plot, the first striking result is that 9/22 of the galaxies with $def_{\rm HI}\geq0.3$ have $def_{\rm H_2}$ values greater than 0.2 dex, while the region characterised by  $def_{\rm HI}<0.3$ and $def_{\rm H_2}\geq0.2$ is almost empty.  This distribution proves the existence of the molecular deficiency in a significant number of galaxies which are HI poor. 
This figure indicates that $def_{\rm H_2}$ and $def_{\rm HI}$ do not follow a tight correlation, but this is not surprising.
In fact, we do not expect effects that deplete the HI to act in the same way on the H$_2$ gas which is more strongly bound. Rather than a correlation, it is evident that galaxies lie in three regions roughly defined by our criterion of isolation and molecular deficiency (dashed lines). Based on Fig. \ref{codef}, we divide our sample in three classes: {\it unperturbed} or  {\it non--deficient} galaxies ($def_{\rm HI}<0.3$ and $def_{\rm H_2}<0.2$),  {\it gas--deficient} galaxies ($def_{\rm HI}\geq0.3$ and $def_{\rm H_2}\geq0.2$) and galaxies which are only {\it HI--deficient} ($def_{\rm HI}\geq0.3$ and $def_{\rm H_2}<0.2$). 
This classification should not be considered strict, since galaxies are distributed continuously rather than in distinct populations, but it is useful for the analysis. We test the robustness of our result by  changing the isolation criterion in the range $def_{\rm HI}=0.1-0.4$ and recomputing $def_{\rm H_2}$ after that a new calibration has been derived. We find that the distribution in Figure \ref{codef} does not change significantly. 

\begin{deluxetable}{l c c c l}
%\rotate
\tablewidth{0pt}
\tablecaption{Completness for the surface density profiles.}
\tablehead{
\colhead{Class}&\colhead{Objects}&\colhead{\% Class \tablenotemark{a}}&\colhead{\% Total \tablenotemark{b}}&\colhead{Symbol}\tablenotemark{c}} 
\startdata
Non--deficient&10&40\%&21\%&Circle\\
Non gas--rich&8&38\%&17\%&Line\\
HI--deficient&5&38\%&11\%&Square\\
Gas--deficient&5&56\%&11\%&Triangle 
\enddata
\tablenotetext{a}{\% Class is defined as the number of galaxies with HI maps
 in each class over the total number of objects in that class.}
\tablenotetext{b}{\% Total is defined as the number of galaxies with HI maps
 in each class over 47, the number of objects in the entire sample.}
\tablenotetext{c}{Symbols used for each class in the plots throughout the paper.}
\label{tab2}
\end{deluxetable}

Since the definition of the molecular gas deficiency relies on  Eq. (\ref{hlco})
which sets the reference mass, the behaviour shown in Fig. \ref{codef} could be an artifact
imposed by this calibration. We further explore the robustness of a H$_2$--deficient population by studying the H$_2$--deficiency with surface density gas profiles, which is independent from  the calibration of $\log M_{\rm H_2,exp}$. 
In Figure \ref{coprof} we show the average molecular gas surface density profiles, in bins of normalized galactocentric radius\footnote{This quantity is well defined under the hypothesis that the stellar disk is not truncated during the perturbation. Clearly this is not the case for tidal interactions. In our sample, NGC 3351 is the galaxy most likely to be affected by tidal stripping. However, we note that a value for $R_{25}$ larger than what we have adopted would shift the surface density profile at lower $R/R_{25}$, enhancing the separation between unperturbed and gas--deficient galaxies.} ($R/R_{25}$).
Not all the galaxies in our sample are imaged in HI. Therefore,  for this kind of analysis, we consider only 20 galaxies (See Table \ref{tab1}). Details on the completeness in each class are provided in Table \ref{tab2}.
By comparing the values for the three classes of galaxies (non--deficient with circles, gas--deficient with triangles and HI--deficient with squares), it is evident that gas--deficient galaxies have the lowest molecular gas column densities, while HI--deficient and non--deficient galaxies exhibit a similar behaviour.
The dashed line in Fig. \ref{coprof} shows the average profile excluding those galaxies with $def_{\rm HI}<-0.3$. The similarity of this profile to that of the full non--deficient sample (circles) indicates 
that the observed difference in H$_2$ content between the three galaxy classes is not an artifact enhanced by the presence of gas--rich galaxies in our sample. Based on an independent method, Fig.  \ref{coprof} confirms the existence of H$_2$--deficient galaxies in $\sim 40 \%$ of the HI poor galaxies and $\sim 20 \%$ of all the galaxies in our sample. As discussed in the next section, these values should be regarded as lower limits.

\begin{figure}
\includegraphics[scale=.40]{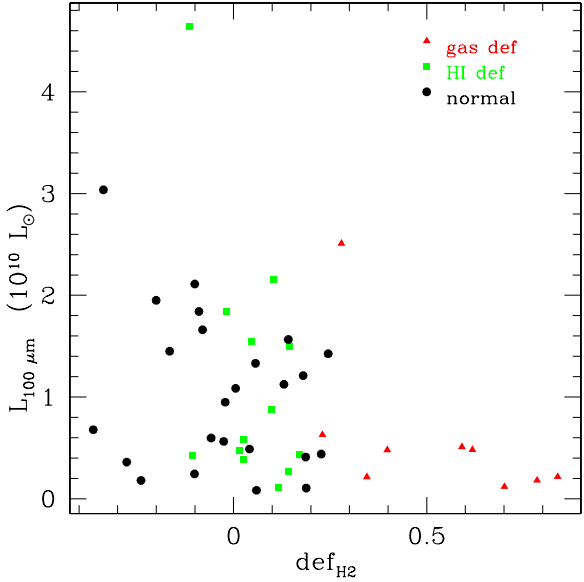}
\caption{Far infrared luminosity at $100 \mu$m as a function of the molecular gas deficiency. Gas--deficient galaxies show the lowest $100\mu$m luminosity.}\label{FIR}
\end{figure}

\section{Discussion}\label{discussion}

\subsection{Comparison with previous studies}\label{subdisc1}
The first issue that we would like to address is why previous studies, based on large CO surveys, did not  reveal a population of H$_2$-deficient galaxies. 
As discussed in the introduction, several studies \citep{sta86,ken86,ken89,cas91,bos97,bos02}
based on large samples both in the Virgo and in the Coma clusters  did not show any evidence for a molecular gas deficiency. 
As discussed by \citet{bos06}, some biases influence the previous works; for example, a constant $X$
conversion factor was often assumed or samples were selected according to FIR selection criteria which prefer CO rich galaxies. Of course, our sample has also been selected to have IRAS fluxes from the Point Source Catalogue higher than 10 Jy at $100\:\mu$m. By comparing the $100\mu$m flux with our $def_{\rm H_2}$
in Figure \ref{FIR}, we notice that gas--deficient galaxies show systematically lower $100\mu$m luminosity; therefore, we might miss a significant fraction of molecular poor galaxies.

Based on this work, we can identify two additional reasons for why previous studies failed to detect the
molecular deficiency: the completeness in the flux sampling and the homogeneity of the data. 
While observing at a single--dish, the integration time required to detect the CO flux is very high (120 minutes at a 12m telescope are needed to reach a sensitivity of 3 mK at a resolution of 15 km/s) and several observations are required to image the entire galaxies. 
It has only recently become possible produce CO images in a reasonable observing time, thanks to the the availability of multibeam receivers that reduce the number of pointings required to fully map the entire disk. Due to these technical limitations, previous studies were based on the observation of one position per galaxy plus one or two positions off-center, along the major axis. The total CO flux  was then recovered by interpolating with an exponential profile, a procedure that inevitably gives only a rough estimate of the proper H$_2$ mass.
By comparing the integrated CO flux from \citet{ken88} and GOLDMine with the Nobeyama and BIMA data, we find a dispersion of up to 1 dex. In contrast, by comparing galaxies in common in the BIMA and Nobayama surveys, the observed discrepancy in the total mass is less than 0.4 dex. Therefore, we believe that the data used in previous studies suffer from  fluctuations larger than the observable H$_2$--deficiency ($< 0.8$ dex in our sample). Another possible problem is related to  the homogeneity of the sample. To increase the statistics, earlier studies often combined CO data from surveys with different sensitivities and diverse techniques to recover the integrated flux, increasing the relative dispersion in the observed H$_2$ mass.
We attribute the fact that  the H$_2$--deficiency has remained hidden in previous analysis to the combination of these two effects. On the contrary, based on high quality and homogeneous  data from the BIMA and Nobeyama surveys,  our work allows for a determination of  molecular gas deficiency.

\subsection{The observed distribution in the H$_2$--deficiency}\label{subdisc2}

The observed distribution in  $def_{\rm H_2}$  deserves a detailed analysis. In particular, we are interested in understanding why HI--deficiency is a necessary  but insufficient condition for H$_2$--deficiency and what the fundamental difference is between galaxies which are only HI--deficient and galaxies which are also molecular gas deficient. The answers to these questions lie in the physics that governs the H$_2$ formation. Molecular hydrogen is synthesized in spiral disks through the interaction of  atomic hydrogen with dust grains, whereas it is destroyed by the UV radiation field. The balance between these processes determines the ratio of molecular to atomic hydrogen. In a series of recent papers,  \citet{kru08I,kru08II} investigate the problem of atomic to molecular ratio with a model that includes the treatment of H$_2$ formation, its self-shielding  and the dust shielding. According to their prescription, the molecular fraction in a galaxy is determined primarily by the product of its hydrogen column density and metallicity. Under the assumption of a two--phase medium \citep{wol95}, the results are largely independent on the UV flux because a change in the UV flux produces a compensating change in the HI density.
Hence, the effects of UV radiation field are largely cancelled out. For the purposes of our analysis, the most important prediction of the model by Krumholz et al. is that the fraction of the gas in molecular form increases as the total gas column density $\Sigma_{gas}=\Sigma_{\rm HI}+\Sigma_{\rm H_2}$ and the metallicity $Z$ increase, providing a large enough shield against the photodissociation.
In agreement with what is observed in local galaxies, the model predicts that the HI column density is expected to saturate at a value of $\sim 10$ M$_\odot$ pc$^{-2}$. Based on the works by Krumholz et al., it is natural to study the behaviour of the total gas column density in our sample to understand the nature of the  H$_2$--deficiency. 

\begin{figure}
\includegraphics[scale=.40]{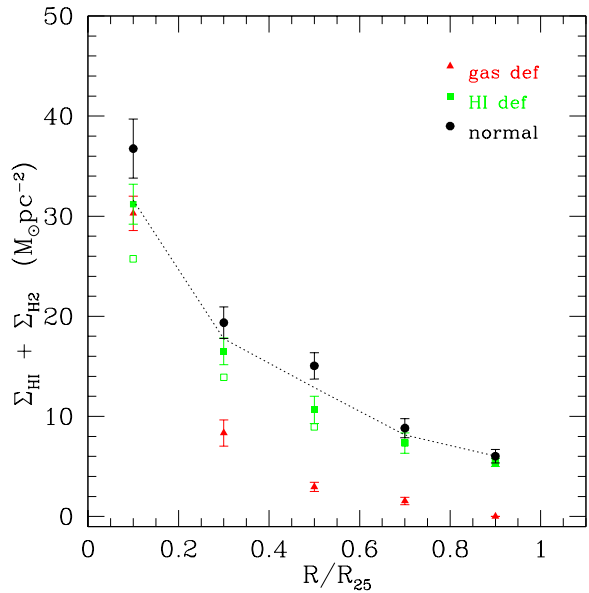}
\caption{Average total gas surface density profiles in bins of normalized radius,
divided according to their $def_{\rm HI}$ and $def_{H_2}$ values. The dashed line
is computed excluding HI rich galaxies with $def_{\rm HI}<-0.3$ and the open squares indicate the profile for HI--deficient galaxies without NGC 4321. Gas--deficient galaxies have in the disk a gas column density below the values required to form molecular hydrogen ($\sim 10$ M$_{\odot}$ pc$^{-2}$).}\label{totgas}
\end{figure}

\begin{figure}
\includegraphics[scale=.40]{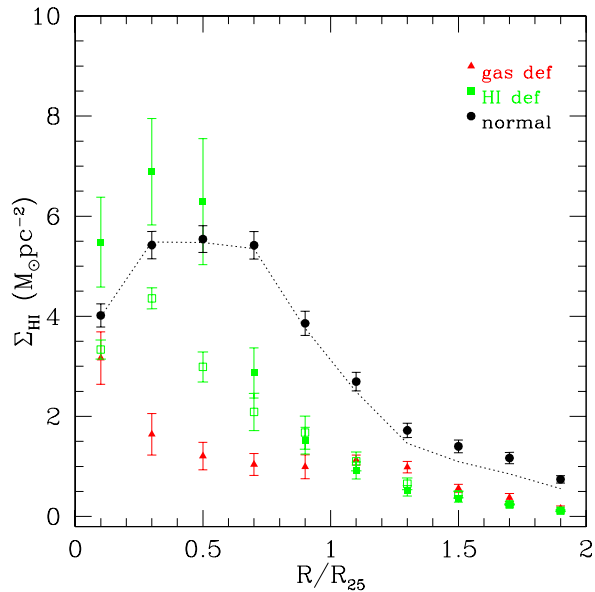}
\caption{Average HI surface density profiles in bins of normalized radius, divided according to their $def_{\rm HI}$ and $def_{H_2}$ values. The dashed line
is computed excluding the HI--rich galaxies with $def_{\rm HI}<-0.3$ and the open squares 
indicates the profile for HI--deficient galaxies without NGC 4321. HI--deficient galaxies have in the inner part of the disk an HI column density large enough to form H$_2$.}\label{hiprof}
\end{figure}

Figure \ref{totgas} shows the average total gas surface density profiles $\Sigma_{\rm HI}+\Sigma_{\rm H_2}$ in bins of the normalized galactocentric radius. Inside the optical radius, the total gas content of the HI--deficient galaxies (squares) is the same as that of the unperturbed galaxies (circles), while gas--deficient galaxies (triangles) have a lower total gas column density. This behaviour reflects the observed atomic and molecular gas content seen in Fig. \ref{codef}, yet add new insights into the nature of the H$_2$ deficiency. Gas--deficient galaxies have a gas column density too low to shield molecules from photodissociation, while HI--deficient galaxies have normal gas content inside their optical radii, which allows the production of molecular hydrogen as in unperturbed galaxies. Indeed,  gas--deficient galaxies lie below the critical shielding column density of $\sim 10$ M$_{\odot}$ pc$^{2}$ while unperturbed and HI--deficient galaxies lie above this threshold. As in Fig. \ref{coprof}, we also plot the average gas profile excluding galaxies with $def_{\rm HI}<-0.3$ (dashed line): the consistency of the profiles with or without HI--rich galaxies excludes any bias in the mean value towards high column density. Inspecting the individual HI profiles, we noticed that NGC 4321 exhibits an unusual high column density. Excluding this object from the analysis (open squares) does not change the results beacause the major contribution to the total gas column density comes from the molecular hydrogen.

At this point we have offered an interpretation for why the HI--deficiency is only a necessary condition for the in H$_2$--deficiency in HI poor galaxies, but a new question arises: if unperturbed and HI--deficient galaxies have a comparable gas column density, why do they have different values of $def_{\rm HI}$? The analysis presented in Figure \ref{hiprof} provides the answer. In this plot we show the average HI surface density profiles in bins of the normalized galactocentric radius up to twice $R_{25}$. 
As for Fig. \ref{totgas}, inspecting the individual HI profiles, we noticed that NGC 4321 exhibits an unusual high column density. To test if this is relevant for our analysis, we over-plot with open squares a new HI profile excluding this galaxy. This profile is no longer fully compatible with the one observed for non--deficient galaxies. It is also evident that the surface density profiles for HI--deficient and gas--deficient galaxies exhibit a different behaviour. The former rises towards the center, approaching the gas column density observed in unperturbed galaxies, while the latter remains at the lowest observed column density at all the radii. Assuming the profile for unperturbed galaxies serves as a template for how the atomic gas is distributed in spiral galaxies, HI--deficient galaxies have a significant residual quantity of gas towards the center that allows the H$_2$ formation,  but have much less HI at larger radii. Therefore, the main contribution to $def_{\rm HI}$ comes from the outer part of the disk where the HI profile rapidly declines. Gas--deficient galaxies  have an HI column density that is reduced relative to that of unperturbed galaxies in both their inner and outer disks and it is only this reduction in the HI content from the inner disk that gives rise to H$_2$ deficiency.
The dashed line in Fig. \ref{hiprof} shows that the observed behaviour is not driven by HI rich galaxies. 

\begin{figure}
\includegraphics[scale=.40]{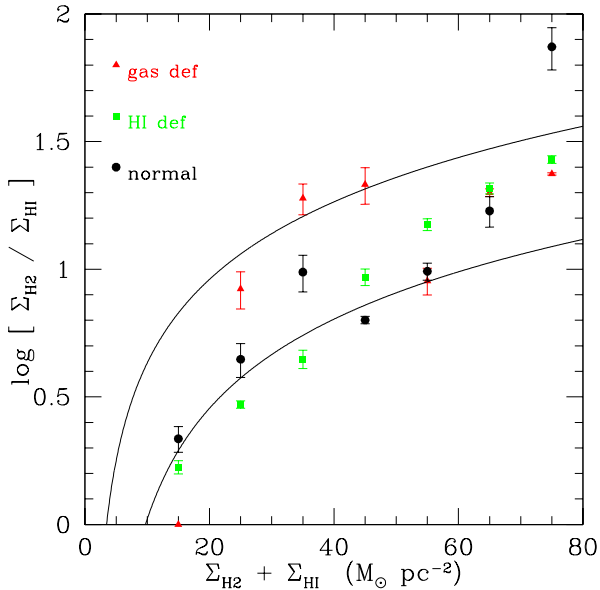}
\caption{Ratio of the molecular to atomic hydrogen as a function of the total gas column density. The solid lines show the expected ratio according to the model by \citet{kru08II} for lower ($[O/H] = - 0.4$) and upper ($[O/H] = 0$) value of metallicity in our sample. At high gas surface densities gas rich and gas poor galaxies exhibit the same behaviour because locally the physics of the H$_2$ formation remains unchanged.}\label{colprof}
\end{figure}

In Figure \ref{colprof} we show the ratio of the molecular to atomic hydrogen as a function of the total gas column density. In this plot, all systematic differences between the classes of galaxy disappear.
Since we are exploring the high density regions, the fact that even gas-deficient galaxies
show the same behaviour observed in non--perturbed and HI--deficient galaxies suggests that
molecule formation is a local process within a galaxy and that the relevant local variable is column density. The solid lines in Fig. \ref{colprof} are the expected values of molecular fraction from the model by  \citet{kru08II}. The model is dependent on the metallicity,  so we consider a lower ($[O/H] = - 0.4$) and an upper ($[O/H] = 0$) value that reasonably correspond to the range of metallicity observed in our sample. There are some approximations that limit the robustness of the comparison between the data and the model. First, the density considered in the model is computed assuming pressure balance,
which may not hold in perturbed galaxies; second, we average the gas column density on physical scales that contain several atomic--molecular complexes, while the model is formulated for a single cloud. Despite all these limitations, there is a satisfactory agreement between data and the model. 
Finally, Figure \ref{colprof} is useful to stress once again the main result of our analysis: 
galaxies which suffer from HI depletion inside the optical disk are on average shifted at lower column density where the production of H$_2$ becomes inefficient. 

\begin{figure*}
\centering
\includegraphics[scale=.50]{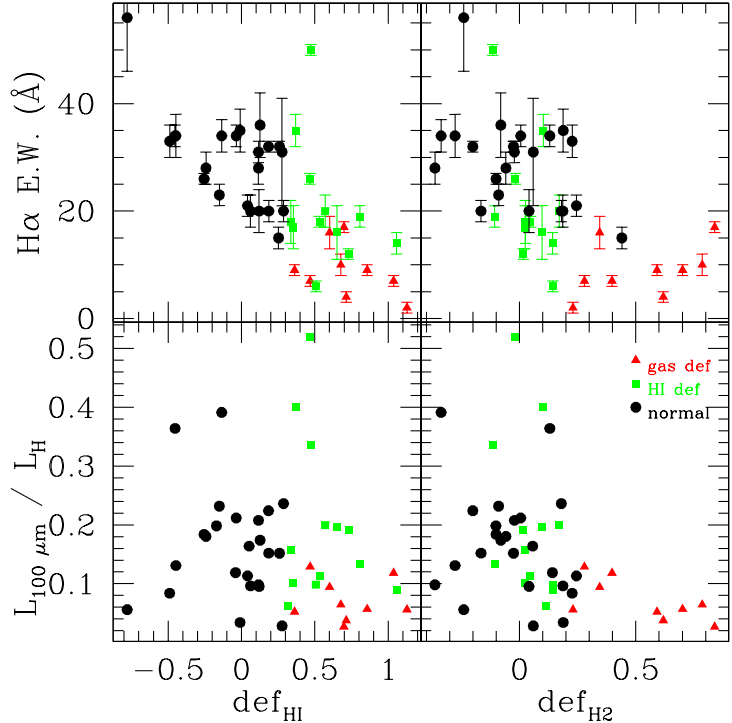}
\caption{Comparison between the atomic and molecular gas deficiency with SFR indicators.
The top panels show the $H\alpha+[NII]$ E.W., while the bottom panels show the $100\mu$m luminosity normalised to the H--band luminosity.  Gas--deficient galaxies have the lowest absolute specific star formation rate as expected if stars form in molecular rich regions.}\label{trittico}
\end{figure*}

\subsection{H$_2$--deficiency and star formation}\label{subdisc3}

Our analysis offers a possible solution to the current inconsistency in studies of the star formation
and galaxy environment. In fact, the observed molecular deficiency is the missing link between the quenching of the star formation rate (SFR) in HI--deficient galaxies \citep[e.g.][]{gav02} and the fact that the bulk of the stars form in molecular regions \citep{won02,ken07,fum08,big08}.
Our simple picture is that, when perturbations reduce the atomic hydrogen column density below the threshold required to produce the molecular hydrogen,  star formation will be suppressed. 

We can test this hypothesis by comparing in Figure \ref{trittico} the integrated atomic and molecular hydrogen content with our indicator for the star formation activity. The top--left panel reproduces the well--known trend \citep[e.g.][]{gav02} between HI--deficiency and the quenching of star formation activity, measured through the $H\alpha+[NII]$  E.W.. Gas--deficient galaxies (triangles) clearly exhibit a lower specific star formation rate than unperturbed (circles) and HI--deficient galaxies (squares). In the top--right panel, we plot the $H\alpha+[NII]$  E.W. versus $def_{\rm H_2}$. Two distinct behaviours
can now be identified: molecular hydrogen deficient galaxies (triangles) have their star formation rate strongly suppressed with respect to molecular rich galaxies (squares and circles). 
Comparing both the top panels, we conclude that the trend between $def_{\rm HI}$  and the star formation rate is in reality a reflection  of the link between  $def_{\rm H_2}$ and  $def_{\rm HI}$ with $def_{\rm H_2}$ and the star formation rate. In more detail, looking at the galaxies with $def_{\rm HI}\geq0.3$ it appears that the HI--deficient galaxies can have star formation rates equal to the values observed in non--deficient galaxies, while gas--deficient galaxies have the lowest $H\alpha+[NII]$  E.W. The dispersion observed in the top--left panel can be interpreted as being primarily a product of the  observed range in the molecular gas deficiency: not all HI--deficient galaxies are molecular deficient and thus not all them have suppressed star formation. This degeneracy is removed by directly comparing the $H\alpha+[NII]$  E.W. with the $def_{\rm H_2}$.
Because we find the same result by comparing the $100\mu$m luminosity normalized to the H--band luminosity with both $def_{\rm HI}$ and $def_{\rm H_2}$ (bottom panels), we are confident that dust extinction or $[NII]$ contamination do not significantly affect our conclusions.

\section{Summary and conclusion}\label{conclusion}
Using high sensitivity and high resolution CO images from the Nobeyama and the BIMA surveys we have studied molecular hydrogen deficiency in spiral galaxies. Our results and conclusions are summarised as follows.
\begin{itemize}
\item[--] Comparing the total masses of molecular and atomic hydrogen,  
we detect H$_2$--deficiency in a subset of galaxies which are HI poor. The same result is found by studying directly the H$_2$ surface density profiles. We observe that molecular gas deficiency is associated with the HI reduction inside the optical disk. The deficiency in the atomic hydrogen is a necessary condition for the molecular gas deficiency, but the $def_{\rm H_2}$ parameter spans a wide range of values even in galaxies which are HI--deficient. In fact, there are HI poor galaxies which have a normal H$_2$ content.
\item[--] When we examine the molecular fraction at a given point in a galaxy as a function 
of the total gas column density, we no longer detect any systematic difference between unperturbed, HI--deficient and gas--deficient galaxies. This suggests that the total gas column density, or something strongly correlated with it, is the primary factor in determining the molecular content of a galaxy. In this picture, a galaxy moves from HI--deficient to gas--deficient when it loses enough gas inside its
optical disk to bring the gas column density below the critical value of $\sim 10$ M$_\odot$ pc$^{-2}$
required for H$_2$ to be present. This behaviour is quantitatively consistent with the models of \citet{kru08I,kru08II}, albeit with significant uncertainties.
\item[--] Since the bulk of the star formation seems connected with the molecular gas, when perturbations reduce the atomic hydrogen column density below the threshold required to produce  molecular hydrogen,  star formation is suppressed. Studying the $H\alpha+[NII]$  E.W. or the $100\mu$m luminosity, we show that  H$_2$--deficient galaxies have less star formation than what is observed in H$_2$--normal galaxies and that the well--known trend between the HI--deficiency and the star formation rate most likely reflects a physical connection between  molecular hydrogen depletion and  star formation activity. 
\end{itemize}

All together, these results corroborate a picture in which the gas phase as a whole feeds the star formation activity. The atomic hydrogen is the essential and primary  fuel for the formation of molecular hydrogen that sustains the bulk of star formation activity. When environmental processes or feedback are able to reduce the atomic gas  inside the optical disk, the HI column density drops below the critical column density required for molecular hydrogen formation and star formation is quenched.
This scenario is supported on cosmological scales, where the HI content in galaxies appears to be unchanged for the past $\sim 10$ Gyr \citep{pro08}. This suggests that the global star formation rate is driven by the accretion of fresh gas from the IGM. 
In addition, our analysis explains how environmental perturbations or galaxy feedback, which remove the atomic gas, can suppress the formation of the H$_2$ and stars. The behaviour seen in the innermost bin ($0\leq R/R_{25}<0.2$) of Figures \ref{coprof}, \ref{totgas} and \ref{hiprof} allows to rule out the hypothesis that AGN feedback is the dominant process responsible for the gas depletion. In fact,  gas is almost unperturbed in the galaxy center. This is expected in our sample where 7/9 of gas--deficient galaxies belong to the Virgo cluster or the Coma 1 cloud in which gas stripping is the most efficient process \citep{bos06,bos06II}.

Ongoing CO surveys as HERACLES by Leroy et al. (in prep) or JCMT, the Nearby Galaxies Legacy Survey by \citet{wil08}, will soon provide larger samples to test our results. A preliminary comparison with the fluxes published by \citet{wil08} appear to support our work. Looking at their Table 4, the four listed galaxies show from the top to the bottom decreasing values of SFR and CO $J:3-2$ luminosity; consistently, we find for them an increasing value of $def_{\rm H_2}$. If confirmed, our results support the argument that the bulk of star formation is connected with the molecular gas phase. Moreover, the ability to derive the molecular gas content starting from an observed HI distribution will provide a useful way to correct the star formation rate expected from local star formation laws \citep{ken98,big08}, including the effect of the molecular gas deficiency which suppresses star formation in rich clusters or in interacting galaxies.

Our analysis leaves open a fundamental question on the implications that  H$_2$--deficiency might have for galaxy evolution. In the standard scenario where star formation is supported by cold gas infall towards the inner part of the galaxy halo \citep{boi99,dek06}, it is fascinating to think that, when the gas flow is halted, the gas is devoured by ongoing star formation. The galaxy depletes its gaseous reservoirs and turns into a passive and quiescent system \citep{vdb76}. This process might explain a first step for the transition between the blue and red sequence in the color--magnitude diagram. Interactions with the intergalactic medium might thus be an alternative process to AGN feedback or galaxy mergers  to explain the departure of star forming systems from the blue sequence. 
In our scenario, two time--scales govern this migration, one set by the atomic gas depletion and the other related to the response of the star formation activity to the gas removal. The first varies according to the physical process responsible for the HI gas removal, but in this phase the galaxy is still forming stars (e.g. Boselli et al 2006). The second is a relatively short time--scale, at most set by the gas consumption through star formation \citep[$\sim 2$ Gyr,][]{big08}\footnote{This value comes from the ratio of gas surface density and star formation rate surface density which is assumed constant. However, gas recycling may increase $\Sigma_{gas}$ up to 40\% and the actual time--scale may be longer.}, but more likely connected with the much shorter life time of the GMCs. This second phase is consistent with a rapid evolutionary path in the color--magnitude diagram. However, although it has been proved that dwarf galaxies can turn into dwarf ellipticals  within $\sim 2$ Gyr  in the local universe \citep{bos08}, it is challenging to prove that massive spirals can turn into giant ellipticals just as a consequence of gas removal through interactions with the cluster IGM. In fact, their angular momentum would be fairly conserved in this kind of interaction. Major mergers as those dominating at higher redshifts must be invoked \citep{fab07}.

\acknowledgments

We thank A. Chung, J. van Gorkom and J. Kenney for their kind permission to use HI maps
from VIVA, prior to publication. M.F. is grateful to B. Devecchi, S. Fabello, S. Colombo and S. Callegari for helpful suggestions and useful discussions.
Support for this work was provided by the National Science Foundation through grant AST-0807739 (to MRK) and by NASA, as part of the Spitzer Space Telescope Theoretical Research Program, through a contract issued by the Jet Propulsion Laboratory, California Institute of Technology (to MRK).
JXP is supported by NSF grant (AST-0709235). 
This research has made use of the GOLDMine Database.
This research has made use of the NASA/IPAC Extragalactic Database (NED) which is operated by the Jet Propulsion Laboratory, California Institute of Technology, under contract with the National Aeronautics and Space Administration.
We acknowledge the usage of the HyperLeda database (http://leda.univ-lyon1.fr).
This research has made use of the SIMBAD database,
operated at CDS, Strasbourg, France.
IRAF is the Image Analysis and Reduction Facility made
available to the astronomical community by the National Optical
Astronomy Observatories, which are operated by AURA, Inc., under
contract with the U.S. National Science Foundation.

\appendix
\section{HI fluxes}\label{appHIfl}

\begin{figure*}
\centering
\includegraphics[scale=.50]{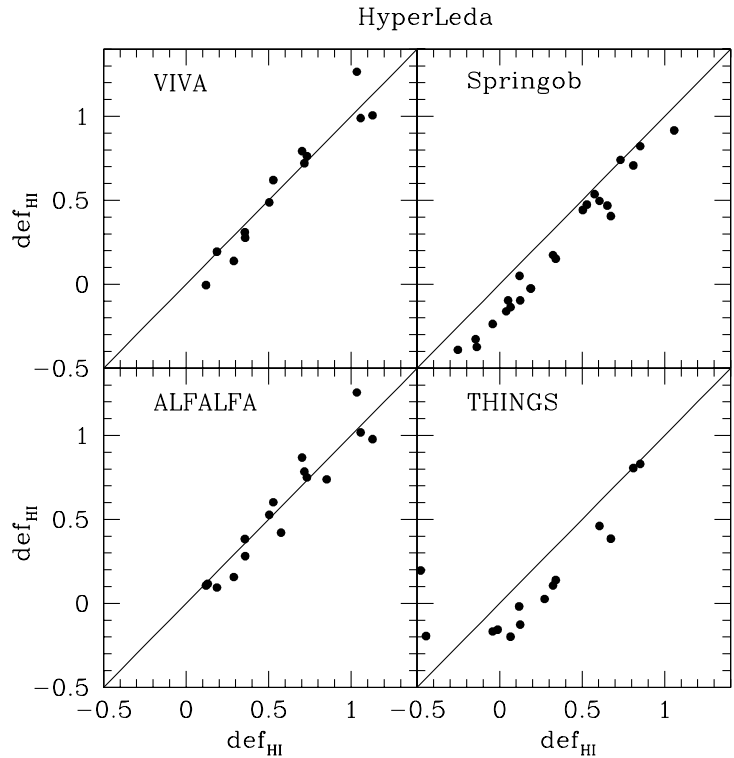}
\caption{Comparison between the HI--deficiency computed using  fluxes from HyperLeda (x--axis) and from THINGS, VIVA, ALFALFA and \citet{spri05} (y--axis). The solid line is the 1:1 correlation.}\label{compdef}
\end{figure*}

As discussed in Sec. \ref{HIdata}, the HI fluxes quoted in the literature for individual galaxies 
are subject to discrepancies which in some case are not negligible. To show that our result
is independent of a particular choice for the HI fluxes, we compute different sets of $def_{\rm HI}$ parameters using fluxes from ALFALFA, THINGS, VIVA and \citet{spri05}. Figure \ref{compdef} shows a comparison of these values with the HI--deficiency computed from HyperLeda fluxes adopted in our analysis.

Inspecting the left--panels, we notice that  $def_{\rm HI}$ computed from ALFALFA and VIVA is in agreement
with our values. Because these two surveys provide HI fluxes in the Virgo cluster where the most HI--deficient objects reside, we conclude that $def_{\rm HI}$ for these galaxies is not affected by important fluctuation. On the contrary, looking at the right--panels, we notice that at low HI--deficiencies a systematic offset appears. Since only few galaxies at low deficiency are available
in ALFALFA or VIVA, an additional comparison is almost impossible and it is difficult to further discuss this difference. However, assuming that HyperLeda fluxes suffer from a systematic offset, we should correct our
$def_{\rm HI}$ towards even lower $def_{\rm HI}$ values and galaxies we classify as non--deficient remain in this class. This is not the case for NGC 5236 and NGC 6946, the two outliers visible in the bottom--right panel. However, these galaxies are more extended than the primary beam mapped by THINGS and we are confident that our value for $def_{\rm HI}$ is more accurate.
  
This comparison confirms that, despite the fluctuation in the $def_{\rm HI}$ parameter, the final distribution for the HI--deficiency and thus our analysis are not significantly affected by a particular choice of the HI flux.

\end{document}